\documentstyle[12pt]{article}
\def\lromn#1{\uppercase\expandafter{\romannumeral#1}}

\begin{document}

\begin{flushright}
TU/98/554\\
RCNS-98-16\\
\end{flushright}

\vspace{1cm}
\begin{center}
\begin{large}

\bf{
Relic Abundance due to Cosmic Pair Annihilation
}

\end{large}

\vspace{36pt}

\begin{large}
Sh. Matsumoto and M. Yoshimura

Department of Physics, Tohoku University\\
Sendai 980-8578 Japan\\
\end{large}

\vspace{4cm}

{\bf ABSTRACT}
\end{center}

Pair annihilation of heavy stable particles that occurs in the
early universe is reconsidered including the off-shell effect
not properly taken into account by the conventional
Boltzmann equation approach.
Our new calculation of the time evolution shows 
that the off-shell effect prolongs
the freeze-out, always with a larger final relic abundance. 
The final yield (number density/temperature$^{\,3}$) is insensitive
to the effective coupling for the annihilation and of order
\( \:
10^{-\,8}\times (M/1\,GeV)^{1/3} \,, 
\: \)
with $M$ the heavy particle mass, if the coupling is not too small.

\newpage

\vspace{0.5cm} 
\hspace*{0.5cm} 
The dark matter problem in cosmology makes it worthwhile to
accurately calculate the relic abundance of weakly interacting
massive particle (WIMP) in modern particle theories.
In most past works 
the annihilation rate has been computed using
a thermally averaged Boltzmann equation \cite{lsp}. 
But estimate based on the on-shell Boltzmann equation
is dubious at very low temperatures, 
and importance of the off-shell effect for the unstable
particle decay has been discussed in \cite{jmy-97-98}.
We extend these works to the annihilation
process such that it can be applied to estimate the relic
abundance of the cold dark matter.

Let us first state our main result and its consequences.
The yield $Y$, the number density per comoving volume, 
is defined relative to (cosmic temperature)$^{3}$;
\( \:
Y = n/T^{3} \,.
\: \)
Introducing the inverse temperature $x = M/T$ for the time
($M$ being the mass of the heavy particle),
the yield follows the evolution equation,
\begin{eqnarray}
&&
\frac{dY}{dx} = -\,\eta (x)\,\left( \,Y^{2} - Y_{{\rm eq}} (x)^{2}\,\right)
\,, 
\label{y evolution} 
\\ &&
\eta (x) = \frac{\langle \sigma _{a}v \rangle\,Mm_{{\rm pl}}}{d\,
x^{2}} \,. 
\end{eqnarray}
The temperature dependent parameter $\eta(x)$ 
is roughly the thermally averaged (on-shell) annihilation rate
$\langle \sigma _{a}v \rangle\,n$ 
divided by the Hubble rate at the temperature $T$,
and is, at $T= M$, 
of order $10^{-3}\,\lambda^{2} \,m_{{\rm pl}}/M$ with
$\lambda $ a small dimensionless coupling and $m_{{\rm pl}}$
the Planck mass.
Here  
\( \:
d = \sqrt{4\pi ^{3}\,N/45} \,, 
\: \)
related to the effective number of particle species $N$ contributing to
the cosmic energy density. 

The equilibrium abundance $Y_{{\rm eq}}(x)$ is usually given by the
thermal number density of the ideal gas form and becomes
at low temperatures
\( \:
Y_{{\rm eq}}(x) \approx (x/2\pi )^{3/2}\,e^{-x}
\: \)
for particles without any conserved quantum number.
This is very small due to the Boltzmann factor $e^{-M/T}$
at $x = M/T \gg  1$.
One may
well question whether this formula is still valid at the low temperature
relevant for the freeze-out.
We shall show below that the off-shell effect modifies this to
\begin{equation}
Y_{{\rm eq}}^{2}(x) = \frac{\delta }{x^{n + 1}} + (\frac{x}{2\pi })^{3}\,
e^{-\,2x} \,.
\end{equation}
The first new term represents a power dependence on the temperature
($\propto T^{n + 1}$), with $\delta $ containing a coupling factor,
for instance $O[10^{-6}] \times $ (4-body coupling)$^{\,2}$, in
a simple model, the boson-pair annihilation model we later explain.
The power $n = 0$ for S-wave annihilation and $n = 1$ for P-wave
annihilation.
In accord with the higher angular momentum $n$, the temperature dependence
of the rate is 
\( \:
\eta (x) = \eta /x^{n + 2} \,.
\: \)

We plot in Fig.1 the solution to the new time evolution equation,
(\ref{y evolution}). We took the parameters relevant to
the boson-pair annihilation model, 
eq.(\ref{parameter for boson-pair model}). 
For comparison a result based on the Boltzmann equation without
the $\delta $ term is also shown.

The analytic estimate of the freeze-out temperature 
\cite{heavy particle decoupling}, important for the relic
abundance, is as follows.
One may consider with a good precision that the yield follows the
equilibrium abundance $Y_{{\rm eq}} $ prior to the freeze-out temperature.
For the simplicity of this estimate we assume that the freeze-out
occurs when the new $\delta /x^{n + 1}$ term dominates over the
exponential one in $Y_{eq}(x)$.
The freeze-out temperature $T_{f}$ is then determined by
\( \:
dY_{{\rm eq}}/dx_{f} = -\,\eta \,Y_{{\rm eq}}^{2}/x_{f}^{n + 2}
\,, 
\: \)
since after this epoch the inverse process is almost frozon,
along with
\( \:
dY/dx = -\,\eta\, Y^{2}/x^{n + 2} \,.
\: \)
Integration of this equation gives the final yield $Y_{0}$
as $T \rightarrow 0$.
When $Y_{f}$ is very small and $x_{f}^{-1} \ll 1$,
$Y_{0} \approx Y_{f}$, as is usually the case.

In terms of the model parameters,
\begin{equation}
\frac{T_{f}}{M} \approx (\frac{n + 1}{2}\,\frac{1}{\eta \,\sqrt{\delta }})
^{\frac{2}{3(n + 1)}} \,, \hspace{0.5cm} 
Y_{f} \approx (\frac{n + 1}{2}\,\frac{\delta }{\eta })^{1/3} \,.
\label{freeze-out for general n} 
\end{equation}
One sees that the freeze-out yield $Y_{f}$ is insensitive to the coupling,
since in simple models the coupling factor cancells in the ratio
\( \:
\delta /\eta \,, 
\: \)
depending on the mass ratio as
\( \:
\delta /\eta \propto M/m_{{\rm pl}} \,.
\: \)
In the boson model
\( \:
Y_{f} \approx 2.4\times 10^{-8}\,N^{1/6}\,(M/1\,GeV)^{1/3} \,.
\: \)
The contribution to the mass density in the present universe is
$\rho _{0} = MY_{f}T_{0}^{3}$ with $T_{0} \approx 3\,K$.
Numerically, 
\begin{equation}
\rho _{0} \approx 4.1 \times 10^{4} \,N^{1/6}\,(M/1\,GeV)^{4/3}
\,eV\,cm^{-3} \,,
\end{equation}
using the ratio $\delta /\eta $ for the boson-pair annihilation model.
Note for comparison that the on-shell Boltzmann equation gives
\( \:
T_{f}/M \approx 1/\ln \eta \,, \hspace{0.2cm} 
Y_{f} \approx (\ln \eta )^{n + 2}/\eta \,.
\: \)

The closure mass density of order 
$1\times  10^{4}\,eV\,cm^{-3}$ thus excludes the WIMP mass range 
far above $1\,GeV$.
In Fig.2 we show the contour map of the present mass density
in the parameter ($\delta /\eta  \,, \sqrt{\delta } $) space,
assuming the mass relation between $\delta /\eta $ and $M/m_{{\rm pl}}$
valid for the boson model.
The parameter region  in which the new
term dominates over the Boltzmann suppressed term in $Y_{{\rm eq}}(x)$
is 
\( \:
0.73\,\delta^{1/3} > (\ln \eta )^{1.6}\,\eta ^{-2/3} \,.
\: \) 

In some parameter region of supersymmetric theories 
$\lambda $ in the boson model is replaced by
\( \:
\approx g^{2}\,(M/\tilde{M})^{2} \,.
\: \)
Here $g$ is a generic gauge coupling constant, and $\tilde{M}$ is
a generic mass of exchanged superparticles for the four-body process, 
usually much larger than the LSP (lightest supersymmetric particle) mass $M$. 
With $g^{2} = 4\pi \alpha \approx 0.1$,
a favored case of $M/\tilde{M} = 0.1$ and $M = O[10\,GeV]$ 
is within the off-shell dominant region given by
\( \:
M < 10^{2}\,GeV \,, 
\: \)
and is excluded by the overclosure from Fig.2.
It is thus wise not to neglect the off-shell efffect
in calculation of the LSP abundance.
Due to the P-wave mixture and detailed numerical factor not known at present,
a realistic calculation in SUSY is called for.

We then turn to a sketch of how the equilibrium abundance $Y_{{\rm eq}}(x)$
above is derived.
A complete derivation involves a fair amount of technical details, and
in this short note we only give the essence of derivation,
relegating technical details and proofs of the statement
to our longer companion paper \cite{my-98-2}.

For simplicity we take an interaction hamiltonian density of
the form, $\frac{\lambda }{2}\,\varphi ^{2}\chi ^{2}$, 
where $\varphi $ is a heavy bosonic
field and $\chi $ a lighter bosonic field.
The dimensionless coupling $\lambda $ must be less than unity for
our result to be a good approximation.
The annihilation channel $\varphi \varphi \rightarrow \chi \chi $ is related
to the scattering channel $\varphi \chi \rightarrow \varphi \chi $ by
a crossing symmetry. 
It is important to recall that a finite time behavior of the quantum
system in thermal medium allows the process such as
\( \:
\chi \leftrightarrow \varphi \varphi \chi  \,, 
\: \)
even if it is kinematically forbidden for the on-shell S-matrix element.

The essential part of our approach is that we separate the system
$\varphi $ part and the environment $\chi $ part, and integrate over the
environment part, assuming a thermal equilibrium for the lighter
particle $\chi $. We thus implicitly assume that $\chi $ has
interaction with other light particles 
or among themselves not written in the above hamiltonian,
to ensure the thermal equilibrium for $\chi $.

Description of this class of the open system
is conveniently done in the path integral approach,
the influence functional method \cite{feynman-vernon}.
The $\chi $ integration is of a Gaussian type,
including the thermal ensemble
average, and one obtains a non-local exponent, symbolically of the form,
\( \:
\varphi ^{2}(1)\,\alpha (1 - 2)\,\varphi ^{2}(2) \,
\: \)
in the influence functional.
The kernel $\alpha $ is known and  complex, reflecting the dissipation
in the open system.
We have developed a new Hartree type of approximation,
replacing four $\varphi $'s here by two $\varphi $'s multiplied by the
correlator $G = i\,\langle \varphi \varphi  \rangle$;
\( \:
\beta (1\,, 2) = -\,i\,\alpha (1 - 2)\,G(1\,, 2) \,.
\: \)
The exponent of this Hartree model 
\( \:
\varphi (1)\beta (1\,, 2)\varphi (2)
\: \)
contains the two-body correlator $G$ to be determined self-consistently.

Derivation of the self-consistency equation is based on a physical 
assumption that
time variation of the particle distribution function occurs
more slowly than indivisual microscopic reactions occur.
The two time argument in the correlator
$\langle \varphi (1)\varphi (2) \rangle$ has then a slowing
varying center of mass (CM) time $(t_{1} + t_{2})/2$, which can be
taken fixed for discussion of the short time dynamics.
A similar separation of the CM and the relative coordinate to distinguish
the slow and the fast process is also taken in other approaches
\cite{real-time th green}, \cite{closed time path}.

As to the short time dynamics 
our reduced Hartree model is equivalent to a Gaussian model that consists
of an infinite set of oscillators \cite{caldeira-leggett 83} 
defined by the given kernel above, once the CM time is fixed.
The oscillator model in this form is exactly solvable and
a convenient form is in \cite{jmy-97-98}.
The time evolution, with regard to the relative time $t_{1} - t_{2}$,
exhibits an exponential decay law, and one may take the infinite
relative time limit which has a non-vanishing equilibrium
contribution for the relevant correlators.
This equilibrium value is then allowed to slowly change
with the CM time $(t_{1} + t_{2})/2$.

The coincident time limit
of two-body correlators needed for our calculation of the occupation number
\begin{equation}
f(\vec{k} ) = \langle \,\frac{\dot{\varphi }_{k}
\dot{\varphi }_{-k}}{2\,\overline{\omega }_{k}}
+ \frac{\overline{\omega }_{k}}{2}\,\varphi _{k}\varphi _{-k}\, 
\rangle - \frac{1}{2}
\end{equation}
(where $\overline{\omega }_{k}$ is a reference energy for the momentum
$\vec{k}$ mode, taken to be the renormalized 
$\tilde{\omega }_{k}(T)$, eq.(\ref{renormalized T-dep mass}),
and $\phi _{k}$ is the Fourier $k$ mode of the $\varphi $ field)
is calculable, for instance, in the operator approach \cite{jmy-97-98}.
The equilibrium occupation number thus calculated
satisfies a self-consistency equation (CM time being omitted here);
\begin{eqnarray}
&&
f_{{\rm eq}}(\vec{k}) + \frac{1}{2} = \frac{1}{4\omega _{k}}\,
\int_{-\infty }^{\infty }\,d\omega \,
\frac{r_{+}(\omega \,, \vec{k})}{(\omega  - 
\tilde{\omega }_{k}(T))^{2}
+ (\pi r_{-}(\omega_{k} \,, \vec{k})/2\omega _{k})^{2}} \,, 
\label{stationary self-c eq1} 
\\ &&
\tilde{\omega }_{k}^{2}(T) 
= \vec{k}^{2} + M_{R}^{2} + \frac{\lambda }{12}\,T^{2}
+ \delta \Pi (\omega _{k} \,, \vec{k}) \,.
\label{renormalized T-dep mass} 
\end{eqnarray}
Here
\( \:
r_{\pm }(\omega \,, \vec{k}) =
r(\omega \,, \vec{k}) \pm r(- \omega \,, \vec{k}) \,, 
\: \)
where ($\beta = 1/T$)
\begin{eqnarray}
&&
r(\omega \,, \vec{k}) =
2\,\int\,\frac{d^{3}k'}{(2\pi )^{3}2\omega _{k'}}\,
\left( \,
\frac{r_{\chi }(\omega  + \omega _{k'} \,, \vec{k} + \vec{k}')}
{1 - e^{-\beta (\omega  + \omega _{k'})}}\,f_{{\rm eq}}(\vec{k'} )
\right.
\nonumber \\ && \hspace*{1.5cm} 
\left.
+ \,
\frac{r_{\chi }(\omega  - \omega _{k'} \,, \vec{k} - \vec{k}')}
{1 - e^{-\beta (\omega  - \omega _{k'})}}\,(1 + f_{{\rm eq}}(\vec{k'})\,) 
\,\right) 
\,,
\end{eqnarray}
is the Fourier transform of the kernel $\beta(x_{1}\,, x_{2})$ 
with respect to the relative coordinate $x_{1} - x_{2}$.
Here $\omega _{k} = \sqrt{k^{2} + M_{R}^{2}}$.
In this self-consistency equation for $f_{{\rm eq}}$
the spectral function $r_{\chi }$ for the two-$\chi $ state
is a given function; it is odd in $k_{0}$ and
\begin{eqnarray}
&& 
r_{\chi }(k _{0} \,, k) = 
\frac{\lambda ^{2}}{16\pi ^{2}}\,\epsilon (k_{0})\,\left( \,
\theta (|k_{0}| - k ) + \frac{2}{\beta k}\,
\ln \frac{1 - e^{-\beta |\omega _{+}|}}
{1 - e^{-\beta |\omega _{-}|}}\,\right)
\,, \label{two-body spectral} 
\end{eqnarray}
with $k = |\vec{k}|$ and
\( \:
\omega _{\pm } = (|k_{0}| \pm k)/2
\: \),
assuming that the $\chi $ particle is massless.
The temperature dependence of the $\varphi $ mass 
is incorporated to $O[\lambda ^{2}]$,
including the finite part of the proper self-energy $\delta \Pi $
after the mass ($M_{R}$ being the renormalized $\varphi $ mass,
hereafter denoted by $M$ for simplicity)
and the wave function renormalization.

The continuous energy integral (\ref{stationary self-c eq1}) 
has a Breit-Wigner form, at the peak of
$\omega = \tilde{\omega }_{k}(T)$ and of width $\Gamma _{k} = \pi r_{-}
(\omega _{k}\,, \vec{k})/\omega _{k}$.
At high temperatures of $T \geq  M$ the entire Breit-Wigner peak region
is integrated, giving the formula in the narrow width limit,
\( \:
f_{{\rm eq}}(\vec{k}) \approx 
r(-\,\omega _{k}\,, \vec{k})/r_{-}(\omega _{k} \,, \vec{k}) \,.
\: \)
At intermediate and low temperatures, a better approximation is
given by a sum of the small resonance term plus a larger continuous
integral;
\begin{eqnarray}
&& \hspace*{0.5cm} 
f_{{\rm eq}}(\vec{k}) = 
\frac{r(-\,\omega _{k}\,, \vec{k})}{r_{-}(\omega _{k} \,, \vec{k})} 
+ \delta \tilde{f}_{{\rm eq}}(\vec{k}) \,, 
\\ && 
\delta 	\tilde{f}_{{\rm eq}}(\vec{k}) = \frac{1}{4\omega _{k}}\,
\int_{-\infty }^{\infty }\,d\omega \,
\,\frac{r_{+}(\omega \,, \vec{k}) - r_{+}(\omega _{k}\,,\vec{k})}
{(\omega - \tilde{\omega }_{k}(T) )^{2} + \Gamma _{k}^{2}/4}
\,. \label{off-shell stationary} 
\end{eqnarray}
The tilded $\delta \tilde{f}_{{\rm eq}}$ 
contains terms to be renormalized away by subtraction,
the renormalized one being denoted by $\delta f_{{\rm eq}}$ 
hereafter.

A straightforward computation gives the off-shell
contribution after renormalization; 
to the leading order of $T/M$ 
\begin{eqnarray}
&& \hspace*{-1cm}
\delta f_{{\rm eq}}(\vec{k}) \approx 
\frac{\zeta (2)\lambda ^{2}}
{16\pi ^{4}}\,\frac{T^{2}}{k\omega _{k}}\,
\int_{0}^{\infty }\,dq\,\frac{q}{2q + \zeta (2)T}\,
\frac{1}{e^{q/2T} - 1}\,\left( \frac{1}{\omega _{k} + \omega _{k - q}}
- \frac{1}{\omega _{k} + \omega _{k + q}}\right) 
\,. \nonumber \\ && 
\end{eqnarray}
Physical processes that contribute to this result
are predominantly inverse annihilation 
$\chi \chi \rightarrow \varphi \varphi $, and 1 to 3 process,
$\chi \rightarrow \chi \varphi \varphi $, which is only
a small fraction of the entire contribution.
The momentum integration of this occupation number gives 
\begin{eqnarray}
&&
\delta n_{{\rm eq}} \approx 
\frac{c\,\lambda ^{2}}{192\pi ^{3}}\,
\frac{T^{4}}{M} \,,
\label{off-shell equil n} 
\\ &&
c = \int_{0}^{\infty }\,dx\,\frac{x^{2}}{4x + \zeta (2)}\,\frac{1}{e^{x} - 1}
\approx 0.270 \,. 
\end{eqnarray}

The time evolution equation may be derived for the occupation
number $f(\vec{k})$. 
This equation is non-Markovian, containing an initial memory term.
A simple Markovian approximation becomes possible, again
under the assumption of the slow time variation of the $\varphi $ number
density.
The Markovian equation thus derived takes the form,
\begin{eqnarray}
&& 
\frac{df(\vec{k} \,, t)}{dt} = -\,\Gamma_{k}
\,\left(\, f(\vec{k} \,, t) - f_{{\rm eq}}(\vec{k} \,, t) \,\right) \,.
\label{kinetic equation} 
\end{eqnarray}

The usual Boltzmann equation follows when one approximates $f_{{\rm eq}}$ 
by dropping the off-shell contribution $\delta f_{{\rm eq}}$.
One may use the explicit formula for $\Gamma _{k}$
(roughly annihilation cross section times target number density $-$
inverse rate), and write the right side of
\( \:
\pi \left( \,r_{-}(\omega _{k} \,, \vec{k} )f(\vec{k})
- r(-\,\omega _{k} \,, \vec{k} )\,\right)
\: \)
in the form equivalent to the Boltzmann equation in the thermal
$\chi $ medium.

Integration of the distribution function over the momentum 
gives a rate equation for the number density which is of
primary interest in the annihilation-scattering problem.
By the momentum integration
the scattering contribution nearly drops out.
This is reasonable, because the scattering does conserve the $\varphi $
particle number, hence the scattering process
does not cause the change of the $\varphi $ number density.

The evolution equation for the number density
is simplified considering the cancellation of scattering terms,
to give
\begin{eqnarray}
&& 
\frac{dn}{dt} = -\,\int\,\frac{d^{3}k}{(2\pi )^{3}}\,
\left( \,
\Gamma _{k}^{{\rm ann}}\,f(\vec{k}\,, t)
- \Gamma _{k}^{{\rm inv}}\,f_{{\rm eq}}(\vec{k} \,, t)  
\,\right) \,,
\end{eqnarray}
where $\Gamma _{k}^{{\rm ann}}\,(\Gamma _{k}^{{\rm inv}})$ is 
the rate keeping the annihilation (inverse annihilation) term.
The first destructive term in the right side
gives the usual, thermally averaged rate
\( \:
\langle \sigma _{a}v \rangle\,n^{2} \,, 
\: \)
while the second production term gives, at low temperatures,
\( \:
\approx 
\langle \sigma _{a}v \rangle\,n_{{\rm th}}\,\delta n_{{\rm eq}}
\,, 
\: \)
with $n_{{\rm th}} = \zeta (3)T^{3}/\pi ^{2}$ and
$\delta n_{{\rm eq}}$ given by (\ref{off-shell equil n}).

In cosmology the thermal environment gradually changes according to
the adiabatic law; the temperature $\propto $ the cosmic scale factor
$1/a(t)$. Along with the number density $\propto 1/a^{3}(t)$,
the time derivative operator is modified to
\( \:
\frac{dn}{dt} + 3H\,n = \cdots \,.
\: \)
Here $H = \dot{a}/a $ is the Hubble parameter.

A detailed behavior of the $\varphi $ number density may be
worked out by examining
\begin{eqnarray}
&&
\frac{dn}{dt} + 3Hn = -\,\langle \sigma _{a}v \rangle
\,(\,n^{2} - \delta \,\frac{T}{M}\,T^{6} - n_{{\rm MB}}^{2}\,) \,, 
\\ && 
\langle \sigma _{a}v \rangle = \frac{\lambda ^{2}}{16\pi M^{2}} \,, 
\hspace{0.5cm} 
\delta  = 0.27\times \frac{\zeta (3)\,\lambda ^{2}}{192\pi ^{5}} \,,
\label{parameter for boson-pair model} 
\end{eqnarray}
where
\( \:
n_{{\rm MB}} \approx (MT/2\pi )^{3/2}\,e^{-M/T} 
\: \)
is the thermal number density of zero chemical potential.
The evolution equation is further simplifed by 
\( \:
\frac{dY}{dt} = 
T^{-3}\,(\,\frac{dn}{dt} + 3H\,n\,)
\,.
\: \)
The equation thus derived is our main result stated 
in the first part of this paper for the S-wave annihilation.
The freeze-out parameters in the boson-pair annihilation model become
\( \:
T_{f}/M \approx 
700\,(N^{1/3}/\lambda ^{2})\,(M/m_{{\rm pl}})^{2/3}
\,, \hspace{0.5cm} 
Y_{f} \approx 0.06 \,N^{1/6}\,(M/m_{{\rm pl}})^{1/3} \,,
\: \)
thus giving the closure density at $M \approx 1 \,GeV$.

\vspace{1cm}
\begin{center}
{\bf Acknowledgment}
\end{center}

This work has been supported in part by the Grand-in-Aid for Science
Research from the Ministry of Education, Science and Culture of Japan,
No. 08640341. The work of Sh. Matsumoto is partially
supported by the Japan Society of the Promotion of Science.


\vspace{1cm}

\begin{Large}
\begin{center}
{\bf Figure caption}
\end{center}
\end{Large}

\vspace{0.5cm} 
\hspace*{-0.5cm}
{\bf Fig.1}

Time evolution for the number density divided by temperature$^{\,3}$.
Parameters are taken from the boson-pair annihilation model in the
text; $n = 0$ and
\( \:
\delta /\eta = 4.6\times 10^{-4}\,\sqrt{N}\,M/m_{{\rm pl}}
\,, \hspace{0.5cm} 
\delta = 5.5 \times 10^{-6}\,\lambda ^{2} \,.
\: \)
The case for $M = 1\,GeV \,, N = 43/4$ 
and indicated values of $\lambda $ is shown, along with
the evolution when the on-shell Boltzmann equation is used.

\vspace{0.5cm} 
\hspace*{-0.5cm}
{\bf Fig.2}

Contour map for the present mass density equated
to the closure $\rho _{c} = 1\times 10^{4}\,
eV\,cm^{-3}$, in the parameter space 
($\delta /\eta  \,, \sqrt{\delta }$), assuming
the relation of the boson model,
\( \:
\delta /\eta = 1.2\times 10^{-22}\,M/GeV 
\: \).
The contours for $0.1\,\rho _{c}$ and $0.01\,\rho _{c}$ are also shown.

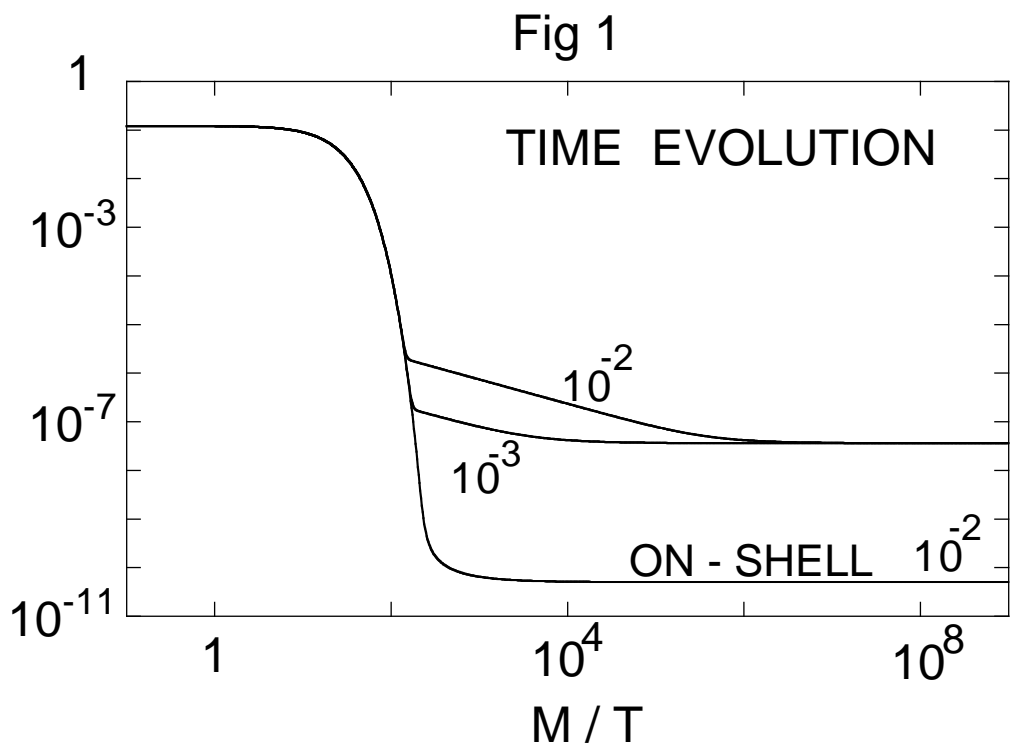
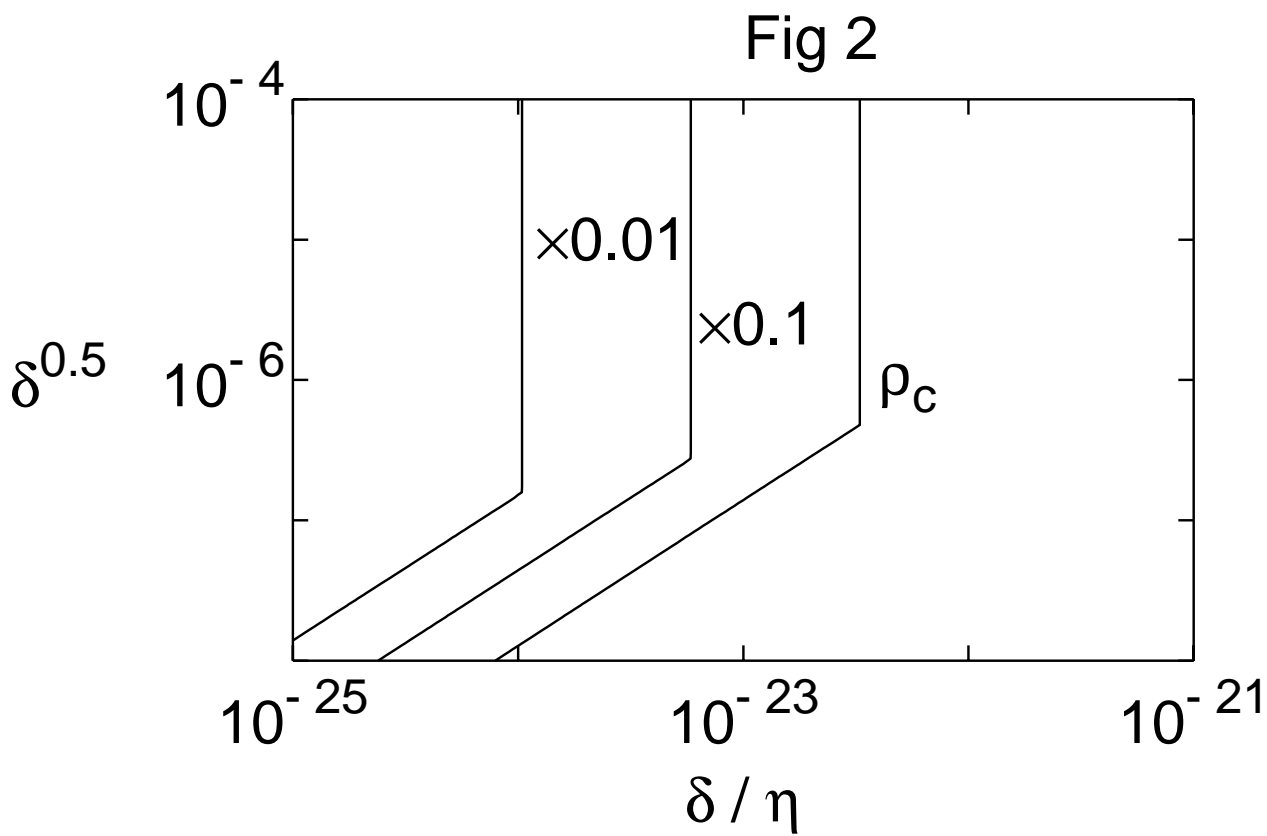
\end{document}